%% file: dstar.tex
\documentclass[12pt,a4paper,dvips]{article}
\usepackage{a4p}
\usepackage{cite,mcite}
\usepackage{graphicx}
\usepackage{physics }
\usepackage{epsfig }
\usepackage{l3_title,ifthen}
\usepackage{Lep}
\date{27 July 1999}
\def\Journal#1#2#3#4{{#1} {\bf #2} (#4) #3}

\def\NPB{{Nucl. Phys.} B}
\def\PLB{{Phys. Lett.}  B}

\def\PRD{{Phys. Rev.} D}
\def\ZPC{{Z. Phys.} C}

 \preprint{99-106}
\newlength{\capwidth}
\def\pb{\mbox{pb$^{-1}$}}

\newcommand{\EE}{\mathrm{e}^+\mathrm{e}^-}

\newcommand{\TT}{\tau^+\tau^-}

\newcommand{\FF}{\mathrm{f} \, \bar{\mathrm{f}}^\prime}

\newcommand{\qq}{\mathrm{q} \, \bar{\mathrm{q}}}
\newcommand{\WW}{\mathrm{W}^+\mathrm{W}^-}

\newcommand{\dzero}{\mathrm{ D^0}}
\newcommand{\pis}{\mathrm{\pi^+_S}}
\newcommand{\dzeropi}{\mathrm{D^0 \,\pi^+_S}}
\newcommand{\dstar}{\mathrm{D}^{* \pm}}
\newcommand{\dstarp}{\mathrm{D}^{*+}}
\newcommand{\dsta}{\mathrm{D}^{*}}
\newcommand{\dstprocess}{\EE \to \EE \dstar \mathrm{X}}

\newcommand{\kpipiz}{\mathrm{K^- \pi^+ \pi^0 }}
\newcommand{\dkpipiz}{\mathrm{D^0 \to K^- \pi^+ \pi^0 }}
\newcommand{\kpi}{\mathrm{K^- \pi^+}}
\newcommand{\etad}{\mathrm{|\eta^{D^*}|}}
\newcommand{\ptd}{p_{T}^{\mathrm{D^*}}}
\begin{document}

\begin{titlepage}
\title{Measurement of Inclusive  $\mathbf{D}^{*\pm}$ Production\\
in Two-Photon Collisions  at LEP}

\author{The L3 Collaboration}


\begin{abstract}

Inclusive production of $\dstar$ mesons in two-photon 
collisions was measured by the L3 experiment at LEP.
The data were collected at a centre-of-mass energy
\mbox{$\sqrt{s} = 189 \GeV{}$} with an integrated 
luminosity of $176.4~\pb$. 
Differential cross sections of the process  $\, \dstprocess \,$  
are determined as  functions of the transverse momentum and
pseudorapidity of the $\dstar$ mesons in the kinematic region
$1 \GeV{} < \ptd < 5 \GeV{}$ and $ \etad < 1.4$. 
The cross section integrated  over this phase space domain is measured to be
$132 \pm 22(stat.) \pm 26(syst.) \; \rm{pb}$.
The differential cross sections are compared with next-to-leading
order perturbative QCD calculations.

\end{abstract}

\submitted

\vfill

\end{titlepage}


\section {Introduction}

The study of charm production in two-photon collisions provides a means
for testing perturbative QCD and for probing the gluon content of the photon \cite{drees}. 
Charmed quarks can be produced in ``direct-photon'' processes, in which the interacting 
photons behave as point-like particles and couple directly to a charmed quark pair.
Another class of processes contributing to the charm production are 
the ``resolved-photon'' processes, where one or both interacting photons fluctuate 
into a flux of partons. In the ``single resolved-photon''  processes the unresolved 
photon interacts with a constituent parton from the resolved photon, whereas in the 
``double resolved-photon'' processes a hard scattering between the  constituent partons
of the two resolved photons takes place. In the next-to-leading order QCD only the sum
of direct and resolved-photon processes is unambiguously defined.
The experimental measurement of differential cross sections for production of open 
charmed particles allows a detailed investigation of the charm production mechanism.

Charm production in two-photon collisions has been measured
at lower centre-of-mass energies at PEP, PETRA, TRISTAN and LEP 
\cite{JADE,TPC-2g,TASSO,TOPAZ,VENUS,AMY,ALEPH}, 
identifying charmed quarks by detecting $\dstar$ mesons, soft pions, inclusive
leptons and $\mathrm{K^{0}_{S}}$ mesons.
In a previous measurement by the L3 experiment \cite{l3cc}, events containing charmed  quarks 
were tagged by detecting electrons and muons from semileptonic decays of charmed hadrons.
In the present study  charmed vector mesons $\mathrm{D^{*}(2010)^{\pm}}$ 
are identified by the small energy released in $\dsta$ decay, applying the mass 
difference technique \cite{mdif} to the decay chains 
\footnote{ The charge conjugate reactions are included
throughout the paper.}

\[\eqalign{~~~~~~~~~~~~~~~~~~~~~~~~~~~~~~~~~~~~~~~~
\dstarp\ra ~ &\dzeropi \cr &\dk \rm{ K^- \pi^+}~~~~~~~~~~~~~~~~~~~~~~~~~~~~~~~~~~~~~~~~~~~~~~~~~~\,(1)& 
\cr &\dk \rm{K^- \pi^+ \pi^0}~~~~~~~~~~~~~~~~~~~~~~~~~~~~~~~~~~~~~~~~~~~~~~~\,(2)}\]

The presence of a low-momentum, ``soft'' pion, $\pis$, ensures that the resolution of 
the mass difference {\mbox{$\mathrm{M(D^0 \pi^+_S) - M(D^0)}$}
is superior to the resolution of the reconstructed  $\dzero$ and $\dstarp$ masses 
themselves. The $\dstarp$ signal appears  as a narrow peak close to
the kinematic threshold in the mass difference distributions
$\mathrm{M(K^- \pi^+ \pi^+_S) - M(K^- \pi^+)}$
and
$\mathrm{M(K^- \pi^+  \pi^0 \pi^+_S) - M(K^- \pi^+  \pi^0)}$.
The combined branching fractions are 
$\mathrm{BR(D^{*+} \to D^0 \pi^+_S) \cdot BR(D^0 \to K^- \pi^+) = 0.0263 \pm 0.0008}$ \\
and
$\mathrm{BR(D^{*+} \to D^0 \pi^+_S) \cdot BR(D^0 \to K^- \pi^+ \pi^0) = 0.0949 \pm 0.0064}$,
as given in Reference \cite{PDG}.


\section {Selection of Hadronic Two-Photon Events}

The data were collected by the L3 detector \cite{L3det} at LEP in 1998
at a centre-of-mass energy \mbox{$\sqrt{s} = 189 \GeV{}$}.
The integrated luminosity is  176.4~$\pb$.

For efficiency studies,  samples of 
$\EE \to  \EE \gamma^* \gamma^* \to \EE  \mathrm{c \bar{c} X}$ 
events are generated using the PYTHIA \cite{pythia} and the JAMVG \cite{verm}
Monte Carlo generators. 
The background sources are simulated by 
JAMVG ($\mathrm{e^{+}e^{-} \ra e^{+}e^{-} \tau^{+} \tau^{-}}$),
KORALZ \cite{koralz} ($\EE \to  \TT(\gamma)$),
KORALW \cite{koralw} ($\EE \to \WW \to \FF \FF$) and
PYTHIA ($\EE \to \qq(\gamma)$, $\mathrm{e^{+}e^{-} \ra e^{+}e^{-} \qq}$ ).
The Monte Carlo events are processed in the same way as the data.

Reconstruction of the decay chains (1) and (2) requires 
a sample of events containing hadronic final states. Events of the type
$\mathrm{\EE \to  \EE \gamma^* \gamma^* \to \EE}  hadrons$ 
are selected by cuts on the energy measured in the electromagnetic
and hadron calorimeters and using tracking information.
To exclude annihilation events, the total visible energy must not exceed $0.4 \, \sqrt{s}$,
the energy deposited in the electromagnetic calorimeter 
must be less than $30 \GeV{}$ and the energy in the hadron calorimeter less than  
$40 \GeV{}$. The  transverse component of the missing momentum vector must be less than
$10 \GeV{}$ and the value of the event thrust must be smaller than 0.95.
Events are required to have at least three  charged particles reconstructed in 
the tracking chamber.

A total of 1253890 events pass the hadron selection cuts. 
The contamination from annihilation processes and two-photon production of tau pairs
is less than $0.5\%$. The subsequent reconstruction, 
which forms  $\dstarp$  candidates from three-prong decays with invariant mass exceeding  
$2 \GeV{}$, suppresses these background contributions to a negligible level.

The trigger efficiency for detecting two-photon hadronic final states is 
$( 87 \pm 3 ) \%$, determined from the data sample itself using a set of independent triggers. 


\section {Mass Reconstruction of  $\mathbf{D}^{*+}$ Decays}

The identification of $\dstarp$ mesons proceeds through two steps: selection of $\dzero$ candidates,
which are then combined with another track to form $\dstarp$ candidates.

Tracks are used for reconstruction of $\dzero$ decays if they satisfy the following 
requirements:

\begin{itemize}
\item
Transverse momentum greater than  $150 \MeV{}$.
\item
At least 40 wire hits measured by the tracking chamber.
\item
Distance of closest approach to the event vertex smaller than 1 mm in the transverse plane.
\end{itemize}

A pair of tracks of opposite charge is required to pass the following
criteria in order to be considered as a $\mathrm{K^- \pi^+}$ system from a $\dzero$ decay: 

\begin{itemize}
\item
The intersection point of the tracks in the transverse plane 
must be displaced by no more than 3 mm away from the event vertex.
\item
$P_\mathrm{K} \cdot  P_\pi > 2 \cdot 10^{-3} $,
where $P_\mathrm{K}$ and $P_\pi $ are the  probabilities,
calculated from the measured energy loss $dE/dX$ of each track,
for kaon and pion mass hypotheses of the corresponding tracks. 
\end{itemize}

The selection of tracks and their combinations into neutral pairs is identical for 
the channels (1) and (2) in order to minimize 
the relative systematic error between the two decay modes.

To reconstruct $\dzero$ decays in the $\kpipiz$ decay mode, a neutral pion is 
added to the selected $\mathrm{K^- \pi^+}$ system. Neutral pion candidates are
formed by a pair of photons, identified as isolated clusters in the
electromagnetic calorimeter, not matched with a charged track.
Photons are accepted for $\pi^0$ reconstruction if they are detected in the barrel 
part of the electromagnetic calorimeter  and their energies are  greater than $100 \MeV{}$.
The $\pi^0$  candidates must have the invariant mass of the photon pair
in the mass window of  $\pm 15 \MeV{}$  around the $\pi^0$  mass.
The decay $\dkpipiz$ proceeds dominantly through one of the 
quasi-two-body intermediate states 
$\mathrm{\bar{K}^{*0} \pi^0}$,
$\mathrm{K^{*-} \pi^+}$ and 
$\mathrm{K^- \rho^+}$  \cite{PDG}.
We require either the invariant mass of a $\mathrm{K\pi}$ subsystem to be within 
$\pm 80 \MeV{}$ of the corresponding  $\mathrm{K^*}(892)$ mass or
the invariant mass of the $\pi^+ \pi^0$ system to be within  $\pm 150 \MeV{}$ 
of the $\rho^+$  mass. If this condition is met for a given intermediate
resonant state, we make use of the P-wave properties of a vector particle decay into
two scalar particles and demand in addition the helicity angle $\theta^*$ of the
corresponding decay cascade to satisfy the condition $|\mathrm{cos} \; \theta^*| > 0.4$. 
The helicity angle $\theta^*$ is defined as the angle between the direction of a decay product
of the vector resonance ($\mathrm{\bar{K}^{*0}}$, $\mathrm{K^{*-}}$ or $\rho^+$) and the 
direction of the pseudoscalar particle ($\pi^0$, $\pi^+$ or $\mathrm{K^-}$) from the $\dzero$ 
decay, calculated in the intermediate resonance rest frame.

To reduce the combinatorial background when reconstructing  $\dzero$ decays
into the $\mathrm{K^- \pi^+}$ final state, the opening angle of the track pair 
in space must be smaller than 2.5 rad.
The combinatorial background for the $\kpipiz$ decay mode is 
suppressed by requiring the solid angle, defined by the directions of flight of the 
three decay products, to be smaller than 2 srad.

The invariant mass of the $\mathrm{K^- \pi^+}$~ system is calculated and 
if it is in the range of  $\pm 100 \MeV{}$  
around the mass of the $\dzero$ mesons \cite{PDG}, the combination is retained as a $\dzero$ 
candidate for the decay channel (1).
The corresponding mass window for candidates in channel (2) is $\pm 50 \MeV{}$.
The different mass windows reflect the corresponding $\dzero$ mass resolutions, as obtained 
by Monte Carlo studies. 
The better resolution of the $\dzero$ reconstruction  in channel (2) is
due to the softer and thus better measured charged particles produced in the three body decay 
and to the use of a well-measured  $\pi^0$.

Finally, the probability that a particular $\mathrm{K^- \pi^+}$ combination comes 
from a $\dzero$ decay in channel (1) is determined from a 1C kinematic fit, 
in which the invariant mass of the pair is constrained to the $\dzero$ mass. 
For the $\kpipiz$ final state we perform a 2C fit,
constraining the mass of the whole system to the $\dzero$ mass and the
two-photon mass to the $\pi^0$ mass. 
A combination is accepted as a $\dzero$ candidate if the confidence level of the fit is 
greater than $0.5 \%$.

In the second  step of the $\dstarp$ reconstruction, we consider all
combinations of a given $\dzero$ candidate with an additional track of positive charge,
assumed to be the soft pion $\pis$, resulting from the $\dstarp$ decay.
A track used as a soft pion candidate must have a transverse momentum greater than  $50 \MeV{}$, 
at least 25 wire hits measured by the tracking chamber, and a distance of closest approach 
to the event vertex smaller than 3 mm in the transverse plane. 

A cut on the transverse momentum of the $\dzeropi$ system, $p_T > 1 \GeV{}$,
is imposed in order to exclude the region of small acceptance of $\dstarp$.

The mass difference distribution $\Delta M = M( \dzeropi ) - M( \dzero )$ 
for the selected $\dzeropi$ combinations in the two channels is shown in Figure~\ref{fig:mdiff}.
The contributions from the decay cascades (1) and (2) accumulate 
in  narrow peaks close to the kinematic threshold. 
The mass difference spectrum is fitted by a sum of 
a Gaussian function for the signal and a term for the background of the form
$\sum_{i=1}^{2} a_i ( \Delta M - m_{\pi})^{b_i}$, where $a_i$ and  $b_i$ are 
free parameters. The peak positions, determined by the fit, are
$145.5 \pm 0.2 \MeV{}$ for the channel (1) and 
$146.1 \pm 0.3 \MeV{}$ for the channel (2),
and agree well with the world average value for the mass difference 
$ m_{\dstarp} - m_{\dzero}$ \cite{PDG}.
The good description of the background by the fit is 
corroborated by a background estimate obtained from the data themselves employing an 
event-mixing technique. 
For this, $\dzero$ candidates from a given event are combined with soft tracks from 
another event, containing  $\dzero$ candidates. The resulting 
background distributions are normalized to the data distributions in the region  
$\Delta M > 0.152 \GeV{}$ and shown in Figure~\ref{fig:mdiff} by the dashed histograms. 
The number of reconstructed $\dstar$ mesons is taken to be the number of observed 
entries in the signal region  $ 0.141 \GeV{} < \Delta M < 0.150 \GeV{} $, less the integral 
of the background fit component over that region.  The $\dstar$ signal is estimated to be  
$102 \pm 17$ events in channel (1) and $42 \pm 11$ in channel (2).

The combinatorial multiplicity in the signal regions  \mbox{$ \Delta M < 0.150 \GeV{}$}
is $1.04 \pm 0.01$ for the  channel (1) and 
   $1.05 \pm 0.02$ for the  channel (2).
There is no overlap of events in this region
between the two channels and since the corresponding peak positions of the $\dstar$ signal 
agree well, we add the two distributions shown in Figure~\ref{fig:mdiff} and 
the resulting mass difference spectrum is shown in Figure~\ref{fig:combdiff}.
The total number of the observed
$\dstar$ mesons, obtained from the fit to the combined spectrum, is $149 \pm 20$. 
If the combined spectrum is split into two distributions for negative and positive charmed 
events, the fit result is $66 \pm 14$  $\mathrm{{D}^{*-}}$ mesons and $76\pm 15$ 
$\dstarp$ mesons.


\section {Inclusive $\mathbf{D}^{*\pm}$ Production Cross Section}

The cross section of inclusive $\dstar$ production (summed over $\mathrm{{D}^{*-}}$ and $\dstarp$)
is determined in the visible kinematic region of experimental acceptance, 
to avoid model-dependent extrapolation uncertainties.
In the present analysis, the selection cuts and the available statistics allow to cover 
the following phase space domain of
$\dstar$   pseudorapidity $\etad$ and transverse momentum  $\ptd$:

\setcounter{equation}{2}
\begin{eqnarray}
 \label{eq:eqn03}
          &\etad  & < 1.4 \\ 
1 \GeV{} <& \ptd  & < 5 \GeV{}. \nonumber
\end{eqnarray}

The differential spectra are obtained by fits to the mass difference distributions
subdivided into three intervals of $\ptd$ or $\etad$,  the other variable being 
integrated over its kinematic region.
Based on Monte Carlo studies, the resolution of the reconstructed $\ptd$ is
determined to be about  $30 \MeV{}$ and the resolution of $\etad$ 
about 0.008 units of pseudorapidity. Thus the smearing and the resulting event migration 
between adjacent bins in the spectra of the reconstructed  $\dstar$ mesons is negligible.

The efficiencies for the reconstruction of $\dstar$ mesons are calculated separately for 
direct-photon processes and  for single resolved-photon processes
with Monte Carlo events generated by the PYTHIA program.  A massive matrix element 
calculation with charmed quark mass value $ m_c = 1.35 \GeV{}$ and the  SaS1d parametrization 
of the parton distributions of the photon \cite{sas1d} was used for the generation
of events.
The reconstruction efficiencies are calculated as a ratio of 
the combined number of reconstructed  $\dstar$ mesons in the two decay channels  to 
the number of generated $\dstar$ mesons and are presented in 
Figure~\ref{fig:eff} as functions of $ \ptd $  and $\etad$.
Evaluated in this way, the efficiencies take into account the corresponding branching 
fractions of the decay modes (1) and (2).
The two sets of efficiencies are similar and  agree within the errors.
This implies that the relative proportion of direct and resolved-photon contributions 
to the charm production 
is not a major source of uncertainty in the determination of the $\dstar$ differential 
cross sections in the phase space region defined by (\ref{eq:eqn03}). 
Equal contributions of both types of charm production 
processes in the kinematic region (\ref{eq:eqn03}) are assumed for the calculation of the 
reconstruction efficiencies,  used for the cross-section evaluation.

The measured cross sections of inclusive $\dstar$ production,
calculated as  functions of $\ptd$  and $\etad$ and integrated over the corresponding bin,
are  listed in Table 1. When evaluating the differential cross sections,
a correction based on the prediction of the 
combined Monte Carlo sample, used for the efficiency determination, is calculated
such as to assign the differential cross sections to the centres of the corresponding bins. 
The values of this correction are found to be different from one only for the 
cross section set $ d \sigma / d \ptd $ in the region of high  $\ptd$.
The  differential cross sections as a function of $\ptd$, obtained after applying 
the bin-centre correction, are listed in Table 1. The differential cross sections assigned 
to the bin centres are plotted in Figure~\ref{fig:xsectpt} and Figure~\ref{fig:xsecteta}.

\begin{table*}[ht]
\begin{center}
\begin{tabular}{|c|c|c||c|c|}
\hline
&&&& \\
$ \ptd \;[\GeV\,]$&$\Delta \sigma_{meas}$ [ pb ]& $ d\,\sigma / d\,\ptd$ [ pb\,/$\GeV \,]$   &
$\etad$ & $\Delta \sigma_{meas}$ [ pb ]  \\
&&&& \\ \hline
1 -- 2 & $ 92.9 \pm 22.2 \pm16.7$  &  $ 92.9\pm 22.2\pm 16.7$& 0.0 -- 0.4& $34.1\pm 8.4 \pm 5.3$  \\ \hline
2 -- 3 & $ 30.1 \pm  8.4 \pm 6.1$  &  $ 28.0\pm  7.8\pm 5.7$ & 0.4 -- 0.8& $47.5\pm 11.0 \pm 9.6$\\ \hline
3 -- 5 & $ 11.3 \pm  3.9 \pm 3.0$  &  $  4.9\pm  1.7\pm 1.3$ & 0.8 -- 1.4& $40.8\pm 15.8 \pm 12.2$\\ \hline
\end{tabular}
\caption{
         Measured cross sections $\Delta \sigma_{meas}$ for inclusive $\dstar $ production, 
         integrated over the corresponding bin.
         The third column of the table gives the differential cross sections as a function of  
         $\ptd$ after bin-centre correction.
         The first errors are statistical, the second systematic.
           }
\end{center}
\end{table*}

The systematic uncertainties on the measured cross sections are estimated by varying the 
selection of tracks and photons and by varying the cuts throughout the  $\dstar$ reconstruction. 
The contribution of the selection procedure to the systematic errors is 
in the range $8\% - 17\%$ affecting mostly the low-$\ptd$ region. The uncertainties in the  
$\kpipiz$ channel are higher than in the $\kpi$ channel.
The  systematic  uncertainties related to the background estimation are determined by using
different forms for the background function in the mass difference fit and by changing the mass 
range of the fit and are found to vary from  $5\%$ to $10\%$.
The $\dstar$ reconstruction efficiencies are calculated also using a Monte Carlo sample 
generated by the JAMVG program which involves only direct processes in charm production,
as well as for PYTHIA generations of direct and resolved processes with varied charmed quark mass value. 
The observed variations of the reconstruction efficiencies are taken into account 
as well as the Monte Carlo statistics, resulting in systematic changes of 
$5\% - 14\%$. The contributions of the various sources of systematic errors are added in quadrature.

The integrated cross section measured in the visible kinematic region is found to be
\begin{center}
\( \sigma(\dstprocess \, ; \;\; 1 \GeV{} < \ptd < 5 \GeV{},\, \etad < 1.4) = 
                132 \pm 22 \pm 26 \; \rm{pb},\)
\end{center}

\noindent
where the first error is statistical and the second systematic.

The integrated cross sections calculated separately for the $\kpi$ and $\kpipiz$ channels,
$\sigma = 124 \pm 24 \; \rm{pb}$ and  $\sigma = 142 \pm 46 \; \rm{pb}$ respectively 
(the errors are statistical only), agree well. This justifies combining the signals 
observed in the two decay modes.

In Figure~\ref{fig:xsectpt} and Figure~\ref{fig:xsecteta} the differential cross sections
are compared to next-to-leading order perturbative QCD computations,
based on a massless approach in calculating  the parton-level cross sections \cite{kniehl}. 
In this scheme the charmed quark is considered to be one of the active flavours inside the photon.
Three different sets of parton density parameterizations of the photon have been used
in the calculations: GS \cite{pdf_gs}, AFG  \cite{pdf_afg} and GRV-HO \cite{pdf_grv}.
The renormalization scale, $\rm \mu_R$, and the factorization scale of the 
photon structure function, $\rm \mu_F$, have been taken as 
$\rm \mu_R =  \mu_F / 2 =  \sqrt{ \it p_{T}^{ \rm 2} +m_c^{ \rm 2}}$
with charmed quark mass value  $ m_c = 1.5 \GeV{}$ \cite{kniehl}.
There is a reasonable agreement between the data and the calculations.
With regard to the variations of the predictions in the region  of low transverse
momentum, we should notice the limited applicability of the massless approach 
for $p_{T} \approx  m_c $ \cite{frix}.


\section {Summary}

The inclusive  production of $\dstar$  mesons in two-photon interactions 
at LEP is measured 
by reconstructing  $\dstarp$  cascade decays involving $\dzero$ decays into
$\kpi$ and $\kpipiz$ final states, as well as  the charge conjugate decay chains. 
The integrated and differential cross sections of inclusive $\dstar$  production 
are determined  in the kinematic region 
$1  \GeV{} < \ptd < 5 \GeV{}$,  $\etad < 1.4$ for which the acceptance
is found to be insensitive to the relative mixture of direct and single resolved-photon 
processes.  In this phase space domain the integrated  cross section is measured to be
$\sigma( \dstprocess ) = 132 \pm 22(stat.) \pm 26(syst.) \; \rm{pb}$.
A reasonable agreement is observed between the measured differential cross sections  and 
the predictions based on  next-to-leading order perturbative QCD calculations.


\section*{Acknowledgments}

We  express our gratitude to the CERN accelerator divisions for the
excellent performance of the LEP machine. We also acknowledge
and appreciate the effort of the engineers, technicians and support staff 
who have participated in the construction and maintenance of this experiment.
We thank B.A.~Kniehl for providing us with the results of QCD cross section calculations.

\clearpage

\input namelist180.tex

\clearpage

  \begin{figure} [p]
  \begin{center}
    \mbox{\epsfig{file=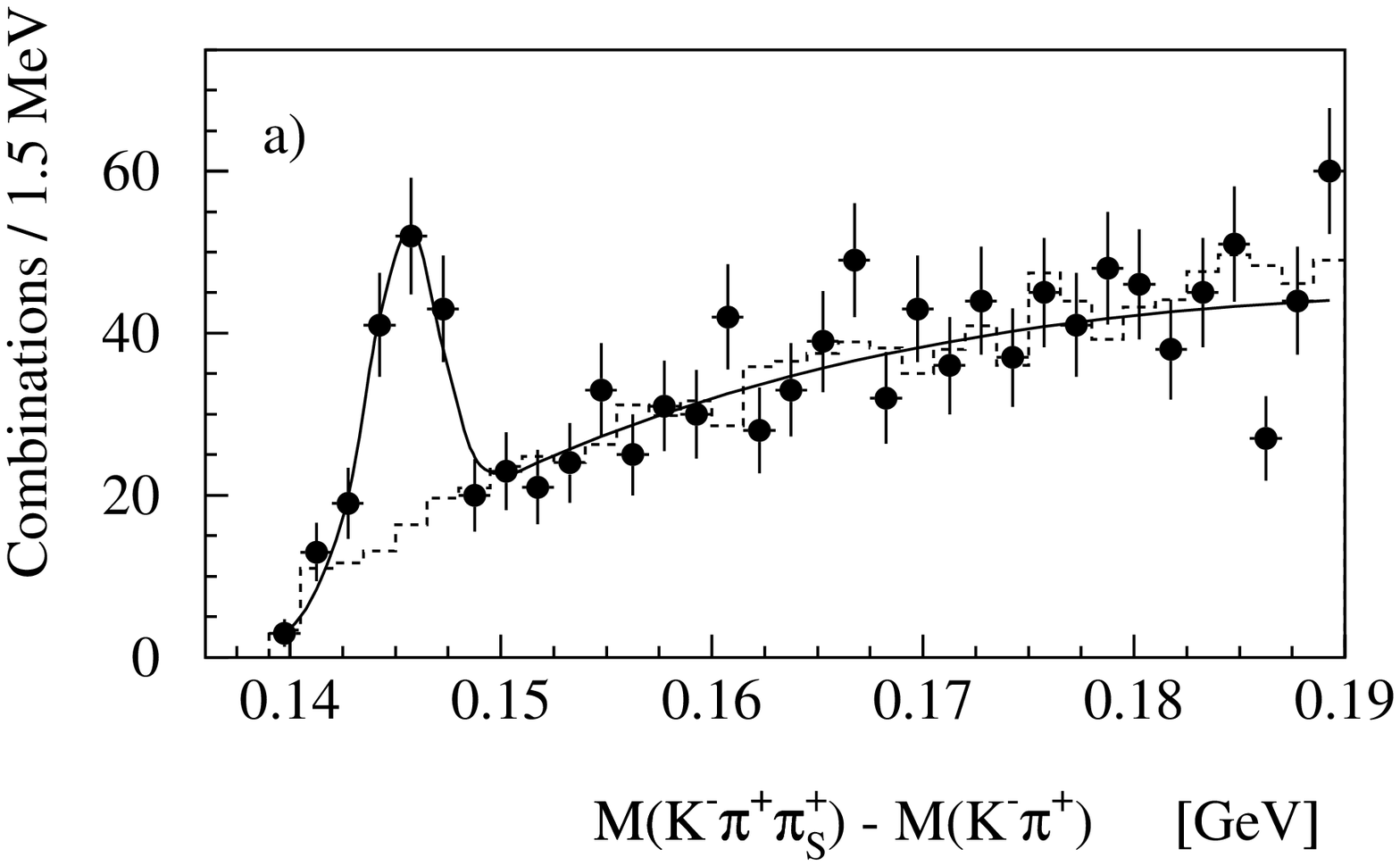,width=0.9\textwidth}}
    \mbox{\epsfig{file=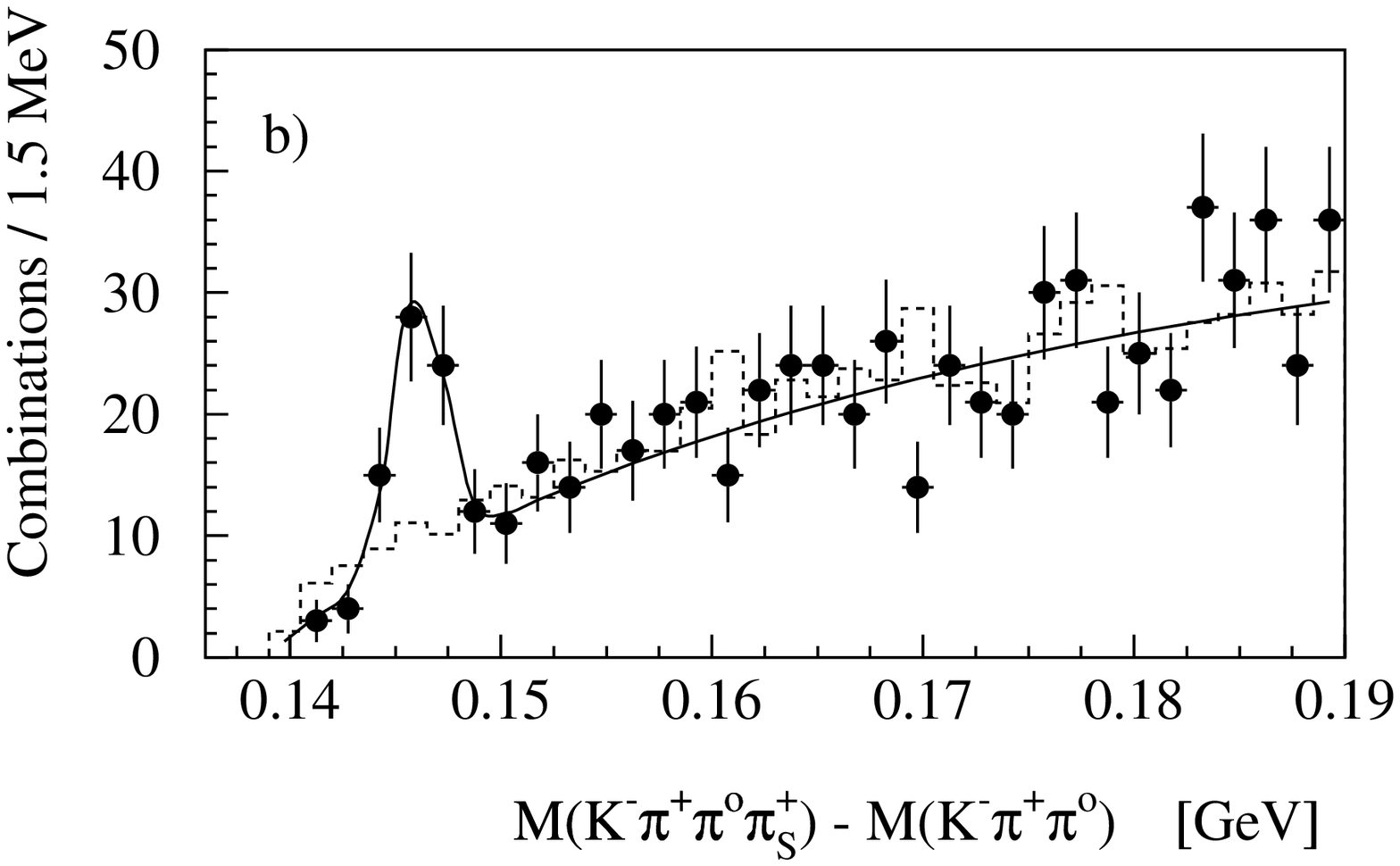,width=0.9\textwidth}}
  \vspace{+4mm}
  \end{center}
  \caption[]{ Mass difference distribution for $\dzero$ decays into
          (a) $\kpi$ and  
          (b) $\kpipiz$.
           The points are data, 
           the line is the result of the fit to the data points 
           used to evaluate the $\dstarp$ signal and 
           the dashed histogram represents a background check, see the text.
           }
\label{fig:mdiff}
\end{figure}
\vfil

\clearpage
  \begin{figure} [ht]
  \begin{center}
    \vspace{+10mm}
    \mbox{\epsfig{file=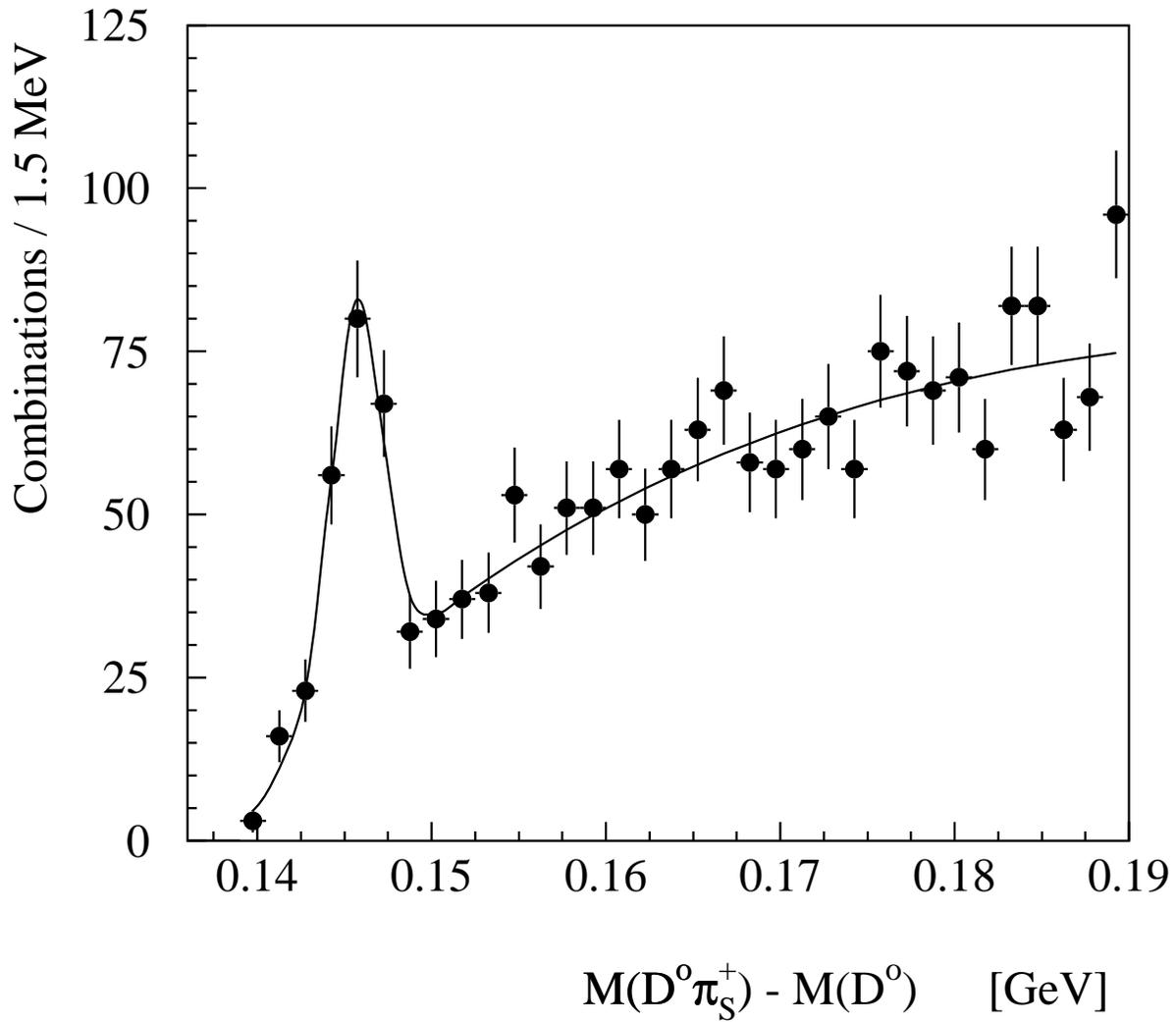,width=1.0\textwidth}}
    \vspace{+10mm}
  \end{center}
  \caption{ Combined mass difference distribution for $\dzero$ decay channels  
           $\kpi$ and $\kpipiz$. The points are data and the line is the result 
           of the fit used to evaluate the $\dstarp$ signal.
           }
\label{fig:combdiff}
\end{figure}
\vfil

\clearpage
  \begin{figure} [p]
  \begin{center}
    \mbox{\epsfig{file=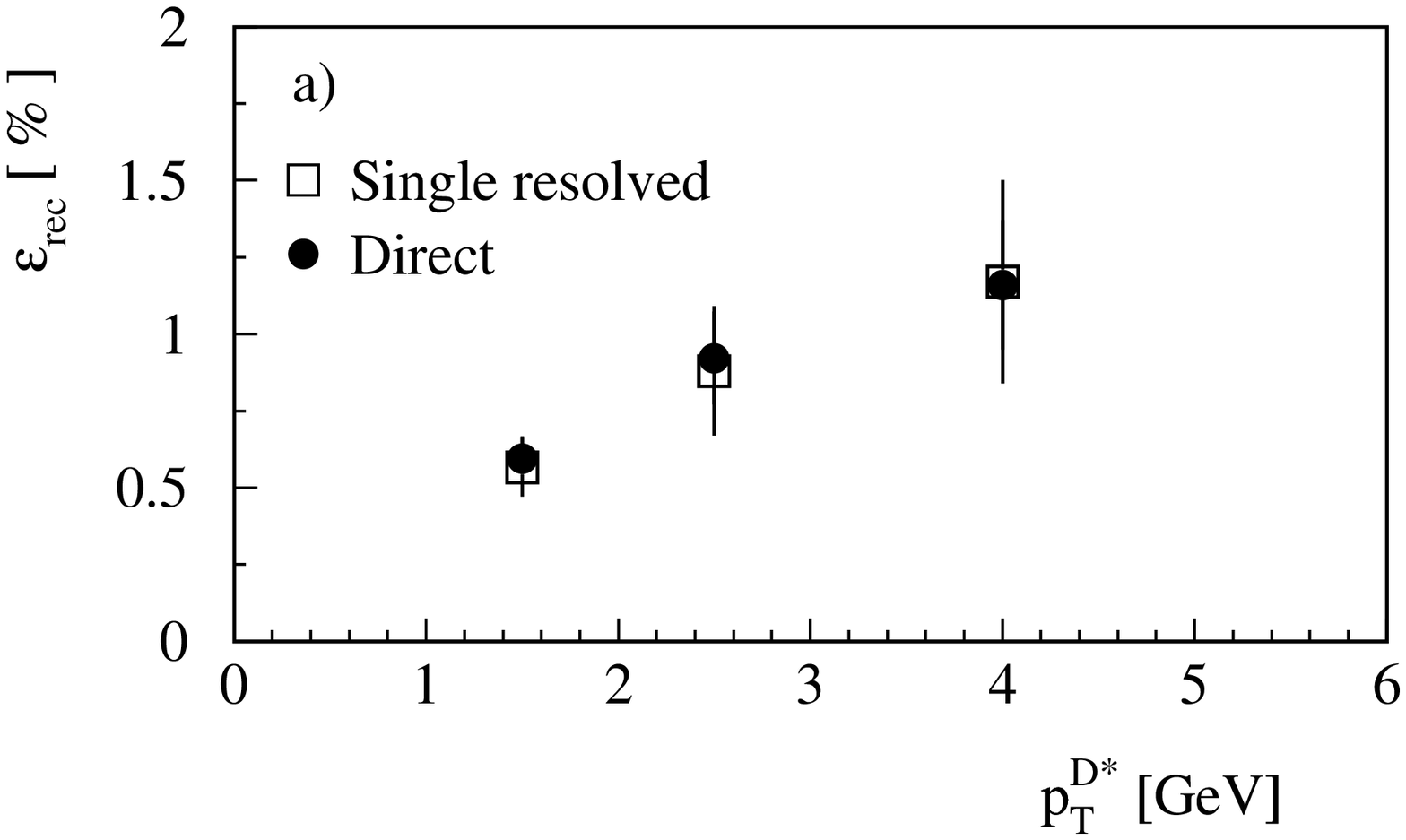,width=0.9\textwidth}}
    \mbox{\epsfig{file=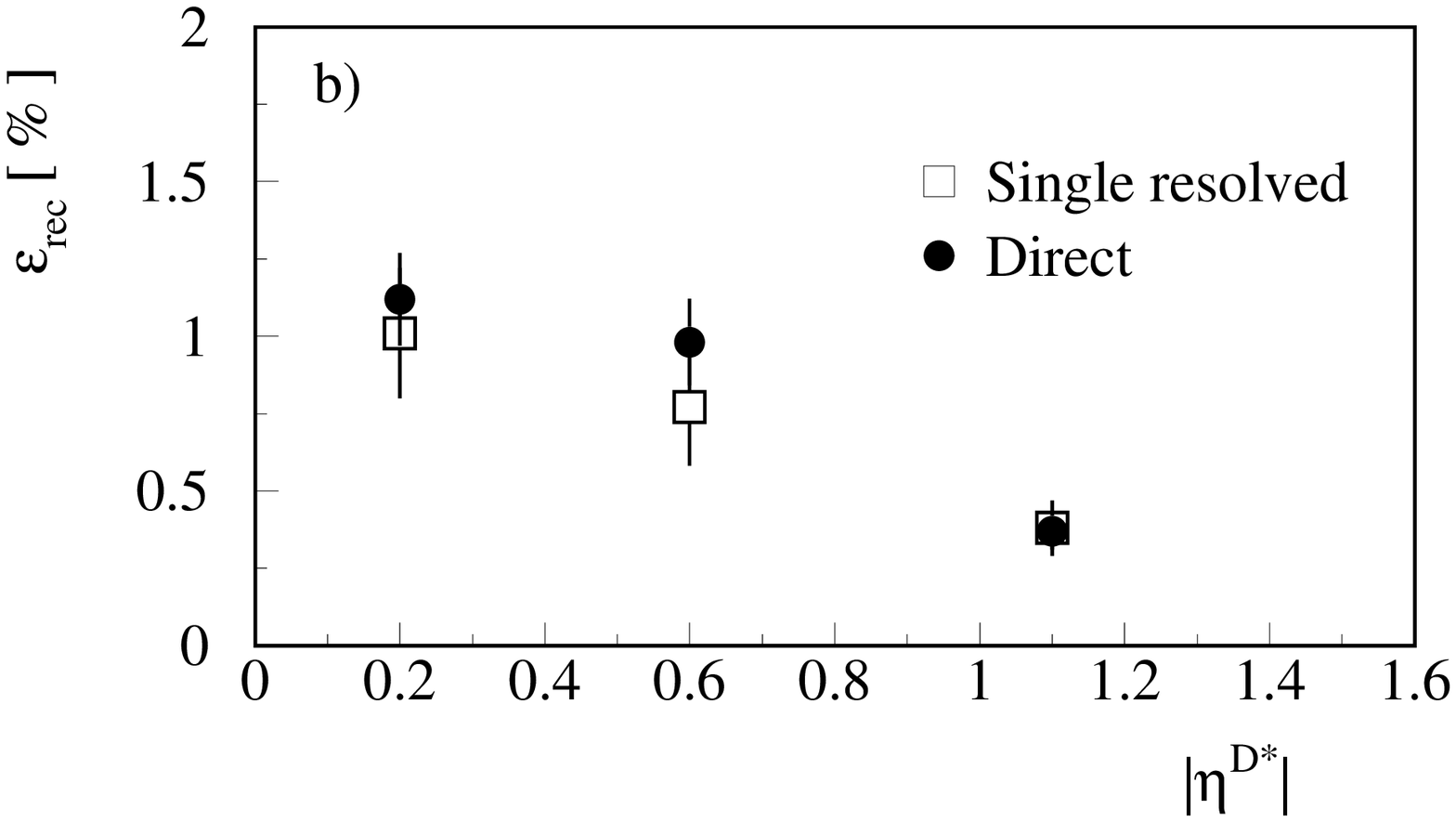,width=0.9\textwidth}}
  \vspace{+4mm}
  \end{center}
  \caption[]{ Reconstruction efficiency of $\dstar$ mesons (including the branching fractions),
            determined from 
            PYTHIA generation of direct and single resolved-photon  processes
          (a) as a function of $\ptd$ for $\etad < 1.4$ and
          (b) as a function of $\etad$ for $1 \GeV{} < \ptd< 5 \GeV{}$.
            }
\label{fig:eff}
\end{figure}
\vfil

\clearpage
  \begin{figure} [ht]
  \begin{center}
    \vspace{+10mm}
    \mbox{\epsfig{file=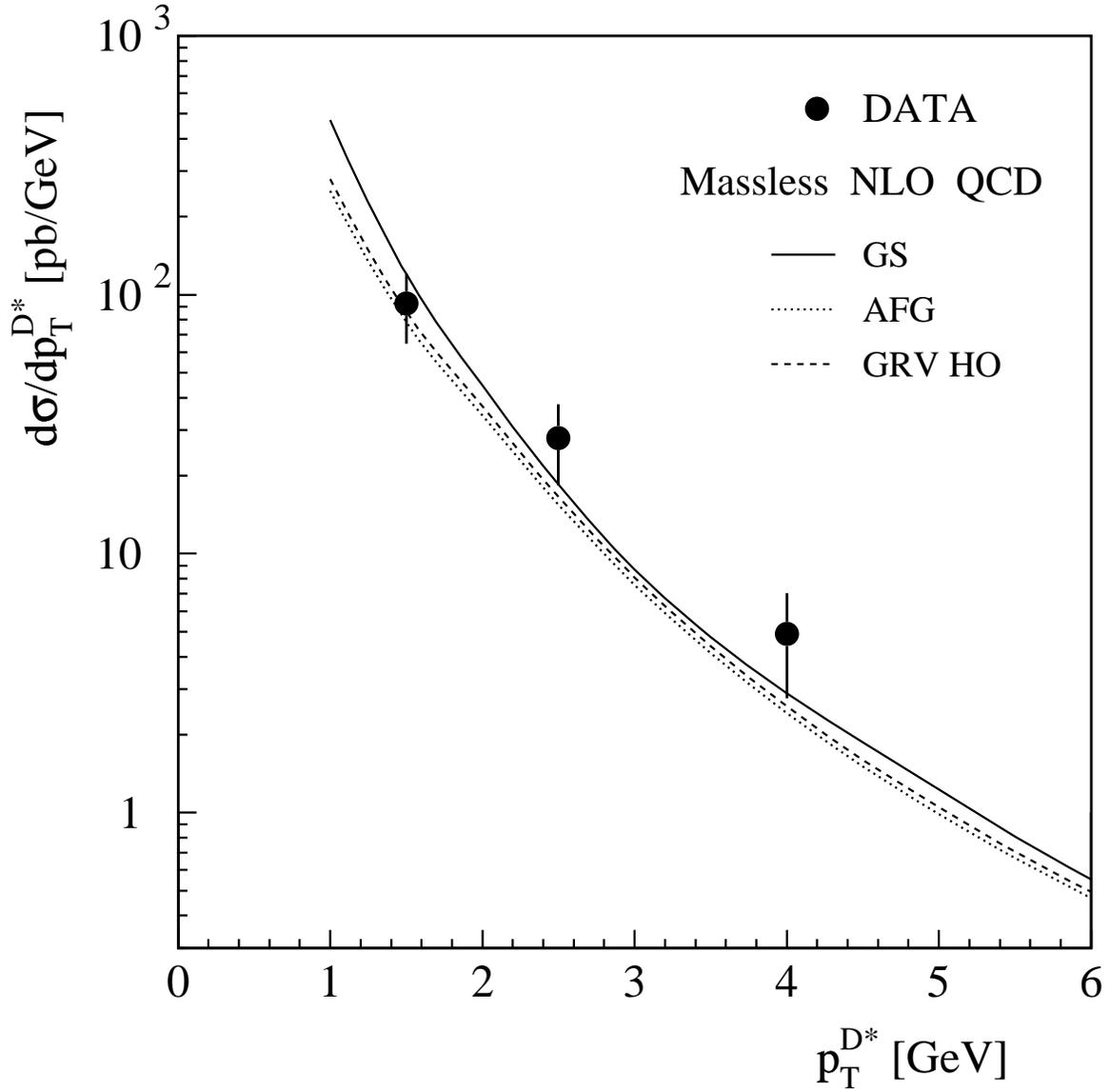,width=1.0\textwidth}}
  \end{center}
  \vspace{+10mm}
  \caption{
           The differential cross section of $\dstar$ production as a function 
           of the transverse momentum of the $\dstar$ mesons for $\etad < 1.4$.
           The points represent the data, the error bars show
           the statistical and systematic errors added in quadrature. 
           The curves represent next-to-leading order QCD calculations 
           \protect\cite{kniehl} for different 
           parameterizations of the parton densities of the photon
           (GS \protect\cite{pdf_gs}, AFG  \protect\cite{pdf_afg} 
           and GRV-HO \protect\cite{pdf_grv}).
           }
\label{fig:xsectpt}
\end{figure}
\vfil

\clearpage
  \begin{figure} [ht]
  \begin{center}
    \vspace{+10mm}
    \mbox{\epsfig{file=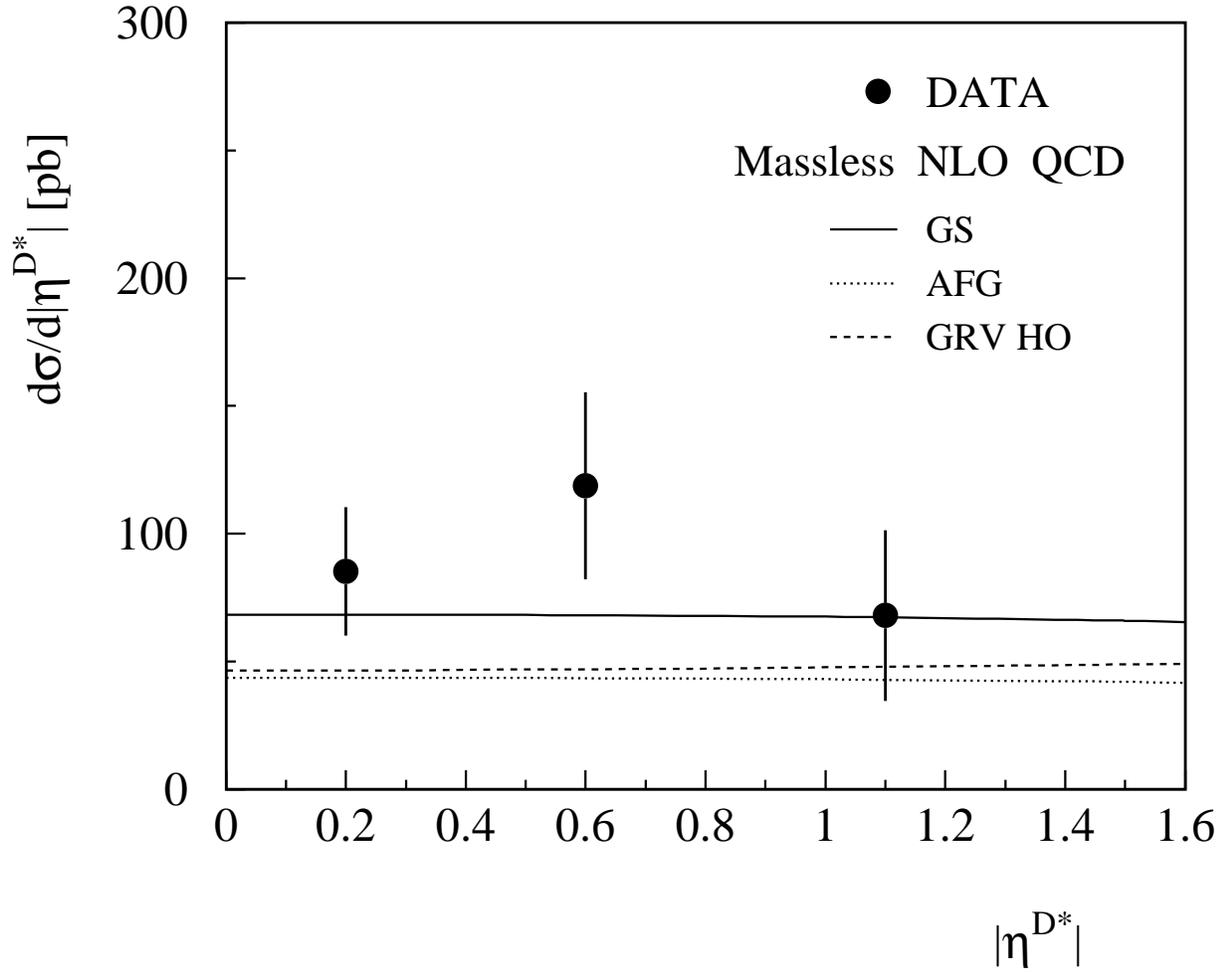,width=1.0\textwidth}}
  \end{center}
  \vspace{+10mm}
  \caption{The differential cross section of $\dstar$ production as a function 
           of the pseudorapidity of the $\dstar$ mesons for $1 \GeV{} < \ptd < 5 \GeV{}$.
           The points represent the data, the error bars show
           the statistical and systematic errors added in quadrature. 
           The curves represent next-to-leading order QCD calculations 
           as in Figure~\ref{fig:xsectpt}.
           }
\label{fig:xsecteta}
\end{figure}
\vfil

\end{document}

%% file: namelist180.tex
\typeout{   }     
\typeout{Using author list for paper 180 -?}
\typeout{$Modified: Fri Jul 23 09:19:07 1999 by clare $}
\typeout{!!!!  This should only be used with document option a4p!!!!}
\typeout{   }
%
%
%
%
%
%

\newcount\tutecount  \tutecount=0
\def\tutenum#1{\global\advance\tutecount by 1 \xdef#1{\the\tutecount}}
\def\tute#1{$^{#1}$}
\tutenum\aachen            
\tutenum\nikhef            
\tutenum\mich              
\tutenum\lapp              
\tutenum\basel             
\tutenum\lsu               
\tutenum\beijing           
\tutenum\berlin            
\tutenum\bologna           
\tutenum\tata              
\tutenum\ne                
\tutenum\bucharest         
\tutenum\budapest          
\tutenum\mit               
\tutenum\debrecen          
\tutenum\florence          
\tutenum\cern              
\tutenum\wl                
\tutenum\geneva            
\tutenum\hefei             
\tutenum\seft              
\tutenum\lausanne          
\tutenum\lecce             
\tutenum\lyon              
\tutenum\madrid            
\tutenum\milan             
\tutenum\moscow            
\tutenum\naples            
\tutenum\cyprus            
\tutenum\nymegen           
\tutenum\caltech           
\tutenum\perugia           
\tutenum\cmu               
\tutenum\prince            
\tutenum\rome              
\tutenum\peters            
\tutenum\salerno           
\tutenum\ucsd              
\tutenum\santiago          
\tutenum\sofia             
\tutenum\korea             
\tutenum\alabama           
\tutenum\utrecht           
\tutenum\purdue            
\tutenum\psinst            
\tutenum\zeuthen           
\tutenum\eth               
\tutenum\hamburg           
\tutenum\taiwan            
\tutenum\tsinghua          
{
\parskip=0pt
\noindent
{\bf The L3 Collaboration:}
\ifx\selectfont\undefined
 \baselineskip=10.8pt
 \baselineskip\baselinestretch\baselineskip
 \normalbaselineskip\baselineskip
 \ixpt
\else
 \fontsize{9}{10.8pt}\selectfont
\fi
\medskip
\tolerance=10000
\hbadness=5000
\raggedright
\hsize=162truemm\hoffset=0mm
\def\r{\rlap,}
\noindent

M.Acciarri\r\tute\milan\
P.Achard\r\tute\geneva\ 
O.Adriani\r\tute{\florence}\ 
M.Aguilar-Benitez\r\tute\madrid\ 
J.Alcaraz\r\tute\madrid\ 
G.Alemanni\r\tute\lausanne\
J.Allaby\r\tute\cern\
A.Aloisio\r\tute\naples\ 
M.G.Alviggi\r\tute\naples\
G.Ambrosi\r\tute\geneva\
% ...
\newpage
\begin{list}{A}{\itemsep=0pt plus 0pt minus 0pt\parsep=0pt plus 0pt minus 0pt
                \topsep=0pt plus 0pt minus 0pt}
\item[\aachen]
 I. Physikalisches Institut, RWTH, D-52056 Aachen, FRG$^{\S}$\\
 III. Physikalisches Institut, RWTH, D-52056 Aachen, FRG$^{\S}$
\item[\nikhef] National Institute for High Energy Physics, NIKHEF, 
     and University of Amsterdam, NL-1009 DB Amsterdam, The Netherlands
\item[\mich] University of Michigan, Ann Arbor, MI 48109, USA
\item[\lapp] Laboratoire d'Annecy-le-Vieux de Physique des Particules, 
     LAPP,IN2P3-CNRS, BP 110, F-74941 Annecy-le-Vieux CEDEX, France
\item[\basel] Institute of Physics, University of Basel, CH-4056 Basel,
     Switzerland
\item[\lsu] Louisiana State University, Baton Rouge, LA 70803, USA
\item[\beijing] Institute of High Energy Physics, IHEP, 
  100039 Beijing, China$^{\triangle}$ 
\item[\berlin] Humboldt University, D-10099 Berlin, FRG$^{\S}$
\item[\bologna] University of Bologna and INFN-Sezione di Bologna, 
     I-40126 Bologna, Italy
\item[\tata] Tata Institute of Fundamental Research, Bombay 400 005, India
\item[\ne] Northeastern University, Boston, MA 02115, USA
\item[\bucharest] Institute of Atomic Physics and University of Bucharest,
     R-76900 Bucharest, Romania
\item[\budapest] Central Research Institute for Physics of the 
     Hungarian Academy of Sciences, H-1525 Budapest 114, Hungary$^{\ddag}$
\item[\mit] Massachusetts Institute of Technology, Cambridge, MA 02139, USA
\item[\debrecen] Lajos Kossuth University-ATOMKI, H-4010 Debrecen, Hungary$^\P$
\item[\florence] INFN Sezione di Firenze and University of Florence, 
     I-50125 Florence, Italy
\item[\cern] European Laboratory for Particle Physics, CERN, 
     CH-1211 Geneva 23, Switzerland
\item[\wl] World Laboratory, FBLJA  Project, CH-1211 Geneva 23, Switzerland
\item[\geneva] University of Geneva, CH-1211 Geneva 4, Switzerland
\item[\hefei] Chinese University of Science and Technology, USTC,
      Hefei, Anhui 230 029, China$^{\triangle}$
\item[\seft] SEFT, Research Institute for High Energy Physics, P.O. Box 9,
      SF-00014 Helsinki, Finland
\item[\lausanne] University of Lausanne, CH-1015 Lausanne, Switzerland
\item[\lecce] INFN-Sezione di Lecce and Universit\'a Degli Studi di Lecce,
     I-73100 Lecce, Italy
\item[\lyon] Institut de Physique Nucl\'eaire de Lyon, 
     IN2P3-CNRS,Universit\'e Claude Bernard, 
     F-69622 Villeurbanne, France
\item[\madrid] Centro de Investigaciones Energ{\'e}ticas, 
     Medioambientales y Tecnolog{\'\i}cas, CIEMAT, E-28040 Madrid,
     Spain${\flat}$ 
\item[\milan] INFN-Sezione di Milano, I-20133 Milan, Italy
\item[\moscow] Institute of Theoretical and Experimental Physics, ITEP, 
     Moscow, Russia
\item[\naples] INFN-Sezione di Napoli and University of Naples, 
     I-80125 Naples, Italy
\item[\cyprus] Department of Natural Sciences, University of Cyprus,
     Nicosia, Cyprus
\item[\nymegen] University of Nijmegen and NIKHEF, 
     NL-6525 ED Nijmegen, The Netherlands
\item[\caltech] California Institute of Technology, Pasadena, CA 91125, USA
\item[\perugia] INFN-Sezione di Perugia and Universit\'a Degli 
     Studi di Perugia, I-06100 Perugia, Italy   
\item[\cmu] Carnegie Mellon University, Pittsburgh, PA 15213, USA
\item[\prince] Princeton University, Princeton, NJ 08544, USA
\item[\rome] INFN-Sezione di Roma and University of Rome, ``La Sapienza",
     I-00185 Rome, Italy
\item[\peters] Nuclear Physics Institute, St. Petersburg, Russia
\item[\salerno] University and INFN, Salerno, I-84100 Salerno, Italy
\item[\ucsd] University of California, San Diego, CA 92093, USA
\item[\santiago] Dept. de Fisica de Particulas Elementales, Univ. de Santiago,
     E-15706 Santiago de Compostela, Spain
\item[\sofia] Bulgarian Academy of Sciences, Central Lab.~of 
     Mechatronics and Instrumentation, BU-1113 Sofia, Bulgaria
\item[\korea] Center for High Energy Physics, Adv.~Inst.~of Sciences
     and Technology, 305-701 Taejon,~Republic~of~{Korea}
\item[\alabama] University of Alabama, Tuscaloosa, AL 35486, USA
\item[\utrecht] Utrecht University and NIKHEF, NL-3584 CB Utrecht, 
     The Netherlands
\item[\purdue] Purdue University, West Lafayette, IN 47907, USA
\item[\psinst] Paul Scherrer Institut, PSI, CH-5232 Villigen, Switzerland
\item[\zeuthen] DESY, D-15738 Zeuthen, 
     FRG
\item[\eth] Eidgen\"ossische Technische Hochschule, ETH Z\"urich,
     CH-8093 Z\"urich, Switzerland
\item[\hamburg] University of Hamburg, D-22761 Hamburg, FRG
\item[\taiwan] National Central University, Chung-Li, Taiwan, China
\item[\tsinghua] Department of Physics, National Tsing Hua University,
      Taiwan, China
\item[\S]  Supported by the German Bundesministerium 
        f\"ur Bildung, Wissenschaft, Forschung und Technologie
\item[\ddag] Supported by the Hungarian OTKA fund under contract
numbers T019181, F023259 and T024011.
\item[\P] Also supported by the Hungarian OTKA fund under contract
  numbers T22238 and T026178.
\item[$\flat$] Supported also by the Comisi\'on Interministerial de Ciencia y 
        Tecnolog{\'\i}a.
\item[$\sharp$] Also supported by CONICET and Universidad Nacional de La Plata,
        CC 67, 1900 La Plata, Argentina.
\item[$\diamondsuit$] Also supported by Panjab University, Chandigarh-160014, 
        India.
\item[$\triangle$] Supported by the National Natural Science
  Foundation of China.
\item[\dag] Deceased.
\end{list}
}
\vfill



